\documentclass[runningheads,a4paper]{llncs}
\usepackage{pbox}
\usepackage{diagbox}
\usepackage{multirow}
\usepackage{amssymb}
\usepackage{graphicx}
\usepackage{epsfig}
\usepackage{amsmath}
\usepackage{amsfonts}
\usepackage{arydshln}
\usepackage{url}
\usepackage{graphicx}
\usepackage{lscape}
\graphicspath{{poster_graphics/}{graphics/}{paper_graphics/}{New_plots/}}
\usepackage{lipsum}
\usepackage{epsfig}
\usepackage{graphicx,epstopdf}
\DeclareGraphicsExtensions{.ps}
\usepackage{verbatim}
\usepackage{multicol}
\usepackage{subfigure}
\usepackage{hyperref}
\hypersetup{
    colorlinks=true,
    linkcolor=blue,
    filecolor=magenta,      
    urlcolor=cyan,
}

\urldef{\mailsa}\path|{alfred.hofmann, ursula.barth, ingrid.haas, frank.holzwarth,|
\urldef{\mailsb}\path|anna.kramer, leonie.kunz, christine.reiss, nicole.sator,|
\urldef{\mailsc}\path|erika.siebert-cole, peter.strasser, lncs}@springer.com|    
\newcommand{\keywords}[1]{\par\addvspace\baselineskip
\noindent\keywordname\enspace\ignorespaces#1}
\begin{document}

\mainmatter  

\title{Deep Learning for Detecting Cyberbullying Across Multiple Social Media Platforms}

\titlerunning{Deep Learning for Detecting Cyberbullying}

%
%
\author{Sweta Agrawal, Amit Awekar}
\authorrunning{Sweta Agrawal, Amit Awekar}

\institute{Indian Institute of Technology, Guwahati\\swetaagrawal20@gmail.com, awekar@iitg.ernet.in}

%
%

\toctitle{Deep Learning for Detecting Cyberbullying Across Multiple Social Media Platforms}
\tocauthor{Author List}
\maketitle

\begin{abstract}

Harassment by cyberbullies is a significant phenomenon on the social media. Existing works for cyberbullying detection have at least one of the following three bottlenecks. First, they target only one particular social media platform (SMP). Second, they address just one topic of cyberbullying. Third, they rely on carefully handcrafted features of the data. We show that deep learning based models can overcome all three bottlenecks.  Knowledge learned by these models on one dataset can be transferred to other datasets. We performed extensive experiments using three real-world datasets: Formspring (\textasciitilde12k posts), Twitter (\textasciitilde16k posts), and  Wikipedia(\textasciitilde100k posts). Our experiments provide several useful insights about cyberbullying detection. To the best of our knowledge, this is the first work that systematically analyzes cyberbullying detection on various topics across multiple SMPs using deep learning based models and transfer learning.

\end{abstract}

%



\section{Introduction}

Cyberbullying has been defined by the National Crime Prevention Council as the use of the Internet, cell phones or other devices to send or post text or images intended to hurt or embarrass another person. Various studies have estimated that between to 10\% to 40\% of internet users are victims of cyberbullying \cite{whittaker15}. Effects of cyberbullying can range from temporary anxiety to suicide\cite{hinduja2010bullying}. Many high profile incidents have emphasized the prevalence of cyberbullying on social media. Most recently in October 2017, a Swedish model Arvida Bystr\"om was cyberbullied to the extent of receiving rape threats after she appeared in an advertisement with hairy legs\footnote{BBC News Article https://goo.gl/t6hQ7c}.

Detection of cyberbullying in social media is a challenging task. Definition of what constitutes cyberbullying is quite subjective. For example, frequent use of swear words might be considered as bullying by the general population. However, for teen oriented social media platforms such as Formspring, this does not necessarily mean bullying (Table~\ref{table:swearanon}). Across multiple SMPs, cyberbullies attack victims on different topics such as race, religion, and gender. Depending on the topic of cyberbullying, vocabulary and perceived meaning of words vary significantly across SMPs. For example, in our experiments we found that for word \lq fat\rq , the most similar words as per Twitter dataset are \lq female\rq\  and \lq woman\rq\ (Table~\ref{tab:meaningchange}). However, other two datasets do not show such particular bias against women. This platform specific semantic similarity between words is a key aspect of cyberbullying detection across SMPs. Style of communication varies significantly across SMPs. For example, Twitter posts are short and lack anonymity. Whereas posts on Q\&A oriented SMPs are long and have option of anonymity (Table~\ref{table:datasets}). Fast evolving words and hashtags in social media make it difficult to detect cyberbullying using swear word list based simple filtering approaches. The option of anonymity in certain social networks also makes it harder to identify cyberbullying as profile and history of the bully might not be available.

Past works on cyberbullying detection have at least one of the following three bottlenecks. First (Bottleneck B1), they target only one particular social media platform. How these methods perform across other SMPs is unknown. Second (Bottleneck B2), they address only one topic of cyberbullying such as racism, and sexism. Depending on the topic, vocabulary and nature of cyberbullying changes. These models are not flexible in accommodating changes in the definition of cyberbullying. Third (Bottleneck B3), they rely on carefully handcrafted features such as swear word list and POS tagging. However, these handcrafted features are not robust against variations in writing style. In contrast to existing bottlenecks, this work targets three different types of social networks (Formspring: a Q\&A forum, Twitter: microblogging, and Wikipedia: collaborative knowledge repository) for three topics of cyberbullying (personal attack, racism, and sexism) without doing any explicit feature engineering by developing deep learning based models along with transfer learning.

We experimented with diverse traditional machine learning models (logistic regression, support vector machine, random forest, naive Bayes) and deep neural network models (CNN, LSTM, BLSTM, BLSTM with Attention) using variety of representation methods for words (bag of character n-gram, bag of word unigram, GloVe embeddings, SSWE embeddings). Summary of our findings and research contributions is as follows.
\begin{itemize}
\item{This the first work that systematically analyzes cyberbullying on various topics across multiple SMPs and applies transfer learning for cyberbullying detection task.}
\item{Presence of swear words is neither necessary nor sufficient for cyberbullying. Robust models for cyberbullying detection should not rely on such handcrafted features.}
\item{Deep Learning based models outperform traditional Machine Learning models for cyberbullying detection task.}
\item{Training datasets for cyberbullying detection contain only a few posts marked as a bullying. This class imbalance problem can be tackled by oversampling the rare class.}
\item{The vocabulary of words used for cyberbullying and their interpretation varies significantly across SMPs.}

\end{itemize}

\section{Datasets}
Please refer to Table~\ref{table:datasets} for summary of datasets used. We performed experiments using large, diverse, manually annotated, and publicly available datasets for cyberbullying detection in social media. We cover three different types of social networks: teen oriented Q\&A forum (Formspring), large microblogging platform (Twitter), and collaborative knowledge repository (Wikipedia talk pages). Each dataset addresses a different topic of cyberbullying. Twitter dataset contains examples of racism and sexism. Wikipedia dataset contains examples of personal attack. However, Formspring dataset is not specifically about any single topic. All three datasets have the problem of class imbalance where posts labeled as cyberbullying are in the minority as compared to neutral posts. Variation in the number of posts across datasets also affects vocabulary size that represents the number of distinct words encountered in the dataset. We measure the size of a post in terms of the number of words in the post. For each dataset, there are only a few posts with large size. We truncate such large posts to the size of post ranked at 95 percentile in that dataset. For example, in Wikipedia dataset, the largest post has 2846 words. However, size of post ranked at 95 percentile in that dataset is only 231. Any post larger than size 231 in Wikipedia dataset will be truncated by considering only first 231 words. This truncation affects only a small minority of posts in each dataset. However, it is required for efficiently training various models in our experiments. Details of each dataset are as follows.

\begin{table*}[tb]
\centering
  \caption{Dataset Statistics}
  \label{tab:datastats}
  \begin{tabular}{|c|c|c|c|c|c|c|}
    \hline
    Dataset & \# Posts & Classes & Length @95\% & Max Length & Vocabulary size  & Source\\
    \hline
    FormSpring	&  12k & 2 & 62 & 1115 & 6058  & \cite{reynolds11}\\
    \hline
    Twitter &  16k & 3 & 26 & 38 & 5653 & \cite{waseem-hovy:2016:N16-2}\\
   \hline
    Wikipedia  & 100k & 2 & 231 & 2846 & 55262  & \cite{wulczyn17}\\	
  \hline
\end{tabular}
\label{table:datasets}
\end{table*}

\underline{Formspring} \cite{reynolds11}: It was a question and answer based website where users could openly invite others to ask and answer questions. The dataset includes 12K annotated question and answer pairs. Each post is manually labeled by three workers. Among these pairs, 825 were labeled as containing cyberbullying content by at least two Amazon Mechanical turk workers. 

\underline{Twitter} \cite{waseem-hovy:2016:N16-2}: This dataset includes 16K annotated tweets. The authors bootstrapped the corpus collection, by performing an initial manual search of common slurs and terms used pertaining to religious, sexual, gender, and ethnic minorities. Of the 16K tweets, 3117 are labeled as sexist, 1937 as racist, and the remaining are marked as neither sexist nor racist. 

\underline{Wikipedia} \cite{wulczyn17}: For each page in Wikipedia, a corresponding talk page maintains the history of discussion among users who participated in its editing. This data set includes over 100k labeled discussion comments from English Wikipedia's talk pages. Each comment was labeled by 10 annotators via Crowdflower on whether it contains a personal attack. There are total 13590 comments labeled as personal attack.

\subsection{Use of Swear Words and Anonymity}
\begin{table*}[t]
\centering
  \caption{Swear Word Use and Anonymity}
  \begin{tabular}{|c|c|c|c|c|c|c|c|c|c|}
    \hline
    Dataset & P(B) & P(S)  & P(A)& P(B$|$S) & P(S$|$B) & P(B$|$A) & P(A$|$B) & P(S$|$A) & P(B$|$ (A\&S)) \\
  \hline
  FormSpring  & 0.06 & 0.16 & 0.53 & 0.22 & 0.59 & 0.08 & 0.71 &  0.20 & 0.25 \\
  \hline
  Twitter & 0.31 & 0.13 & - & 0.42 & 0.18 & - &  - & - & -\\
  \hline
  Wikipedia & 0.12 & 0.17 & 0.27 & 0.49 & 0.69 & 0.25 & 0.56  & 0.27 & 0.65\\
  \hline
  \end{tabular}
\label{table:swearanon}
\end{table*}

Please refer to Table~\ref{table:swearanon}. We use the following short forms in this section: B=Bullying, S=Swearing, A=Anonymous. Some of the values for Twitter dataset are undefined as Twitter does not allow anonymous postings. Use of swear words has been repeatedly linked to cyberbullying. However, preliminary analysis of datasets reveals that depending on swear word usage can neither lead to high precision nor high recall for cyberbullying detection. Swear word list based methods will have low precision as P(B$|$S) is not close to 1. In fact, for teen oriented social network Formspring, 78\% of the swearing posts are non-bullying. Swear words based filtering will be irritating to the users in such SMPs where swear words are used casually. Swear word list based methods will also have a low recall as P(S$|$B) is not close to 1. For Twitter dataset, 82\% of bullying posts do not use any swear words. Such passive-aggressive cyberbullying will go undetected with swear word list based methods. Anonymity is another clue that is used for detecting cyberbullying as bully might prefer to hide its identity. Anonymity definitely leads to increased use of swear words (P(S$|$A) $\geq$ P(S)) and cyberbullying (P(B$|$A)$\geq$P(B), and P(B$|($A\&S))$\geq$P(B)). However, significant fraction of anonymous posts are non-bullying (P(B$|$A) not close to 1) and many of bullying posts are not anonymous (P(A$|$B) not close to 1).  Further, anonymity might not be allowed by many SMPs such as Twitter.

\section{Related Work}
Cyberbullying is recognized as a phenomenon at least since 2003 \cite{servance03}. Use of social media exploded with launching of multiple platforms such as Wikipedia (2001), MySpace (2003), Orkut (2004), Facebook (2004), and Twitter (2005). By 2006, researchers had pointed that cyberbullying was as serious phenomenon as offline bullying \cite{patchin06}. However, automatic detection of cyberbullying was addressed only since 2009 \cite{yin09}. As a research topic, cyberbullying detection is a text classification problem. Most of the existing works fit in the following template: get training dataset from single SMP, engineer variety of features with certain style of cyberbullying as the target, apply a few traditional machine learning methods, and evaluate success in terms of measures such as F1 score and accuracy. These works heavily rely on handcrafted features such as use of swear words. These methods tend to have low precision for cyberbullying detection as handcrafted features are not robust against variations in bullying style across SMPs and bullying topics. Only recently, deep learning has been applied for cyberbullying detection \cite{badjatiya17}. Table~\ref{tab:papers} summarizes  important related work.

\section{Deep Neural Network (DNN) Based Models}
We experimented with four DNN based models for cyberbullying detection: CNN, LSTM, BLSTM, and BLSTM with attention. These models are listed in the increasing complexity of their neural architecture and amount of information used by these models. Please refer to Figure 1 for general architecture that we have used across four models. Various models differ only in the Neural Architecture layer while having identical rest of the layers. CNNs are providing state-of-the-results on extracting contextual feature for classification tasks in images, videos, audios, and text. Recently, CNNs were used for sentiment classification \cite{kim2014convolutional}. Long Short Term Memory networks are a special kind of RNN, capable of learning long-term dependencies. Their ability to use their internal memory to process arbitrary sequences of inputs has been found to be effective for text classification\cite{johnson2016supervised}. Bidirectional LSTMs\cite{zhou2016text} further increase the amount of input information available to the network by encoding information in both forward and backward direction. By using two directions, input information from both the past and future of the current time frame can be used. Attention mechanisms allow for a more direct dependence between the state of the model at different points in time. Importantly, attention mechanism lets the model learn what to attend to based on the input sentence and what it has produced so far.

 \begin{figure}[tb]
\centering
\includegraphics[width=12cm ,height=2cm]{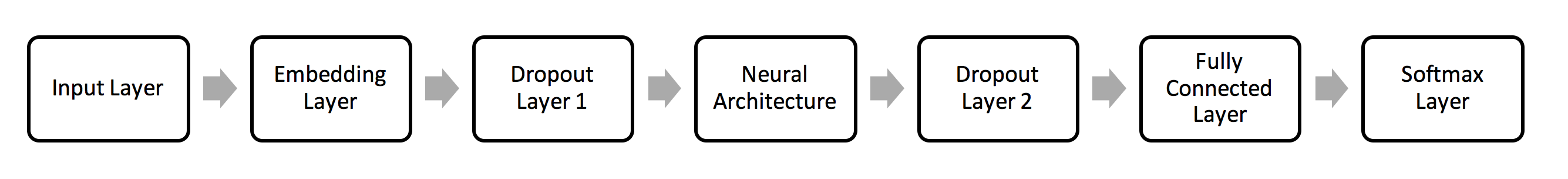}
\caption{Model Architecture}
\end{figure}

The embedding layer processes a fixed size sequence of words. Each word is represented as a real-valued vector, also known as word embeddings. We have experimented with three methods for initializing word embeddings: random, GloVe\cite{pennington2014glove}, and SSWE\cite{tang2014learning}. During the training, model improves upon the initial word embeddings to learn task specific word embeddings. We have observed that these task specific word embeddings capture the SMP specific and topic specific style of cyberbullying. Using GloVe vectors over random vector initialization has been reported to improve performance for some NLP tasks. Most of the word embedding methods such as GloVe, consider only syntactic context of the word while ignoring the sentiment conveyed by the text. SSWE method overcomes this problem by incorporating the text sentiment as one of the parameters for word embedding generation. We experimented with various dimension size for word embeddings. Experimental results reported here are with dimension size as 50. There was no significant variation in results with dimension size ranging from 30 to 200.

To avoid overfitting, we used two dropout layers, one before the neural architecture layer and one after, with dropout rates of 0.25 and 0.5 respectively. Fully connected layer is a dense output layer with the number of neurons equal to the number of classes, followed by softmax layer that provides softmax activation. All our models are trained using backpropagation. The optimizer used for training is Adam and the loss function is categorical cross-entropy. Besides learning the network weights, these methods also learn task-specific word embeddings tuned towards the bullying labels (See Section~\ref{subsec:taskword}). Our code is available at: \url{https://github.com/sweta20/Detecting-Cyberbullying-Across-SMPs}.

\section{Experiments}
Existing works have heavily relied on traditional machine learning models for cyberbullying detection. However, they do not study the performance of these models across multiple SMPs. We experimented with four models: logistic regression (LR), support vector machine (SVM), random forest (RF), and naive Bayes (NB), as these are used in previous works (Table~\ref{tab:papers}). We used two data representation methods: character n-gram and word unigram. Past work in the domain of detecting abusive language have showed that simple n-gram features are more powerful than linguistic and syntactic features, hand-engineered lexicons, and word and paragraph embeddings\cite{nobata16}. As compared to DNN models, performance of all four traditional machine learning models was significantly lower. Please refer to Table~\ref{tab:traditionalML}. 

\begin{table*}[tb]
\centering
  \caption{Results for Traditional ML Models Using F1 Score}
  \label{tab:traditionalML}
  \begin{tabular}{|c|c|c|c|c|c|c|c|c|c|}
    \hline
    Dataset &label & \multicolumn{4}{|c|}{Character n-grams} &\multicolumn{4}{|c|}{Word unigrams} \\
    \cline{3-10}
     & &  LR & SVM & RF &NB &LR & SVM & RF & NB\\
  \hline
    Formspring & bully &  0.448 & 0.422 & 0.298& 0.359 & 0.489&0.463 & 0.264&0.025\\
    \hline
    Twitter& racism &0.723&0.676 & 0.752& 0.686 &0.738 &0.772&0.739&0.617 \\
   & sexism &0.729&0.688 &0.720 & 0.647 &0.762 &0.758&0.755 &  0.635  \\
    \hline
    Wiki& Attack &0.694 &0.677 &0.674 & 0.655& 0.711 &0.686&0.730 & 0.659\\
\hline
\end{tabular}
\end{table*}

All DNN models reported here were implemented using Keras. We pre-process the data, subjecting it to standard operations of removal of stop words, punctuation marks and lowercasing, before annotating it to assigning respective labels to each comment. For each trained model, we report its performance after doing five-fold cross-validation. We use following short forms.
\begin{itemize}
\item{Datasets: F (Formspring), T (Twitter), W (Wikipedia) }
\item{Datasets with oversampling of bullying posts: F+ (Formspring), T+ (Twitter), W+ (Wikipedia) }
\item{Evaluation measures: P (Precision), R (Recall), F1 (F1 score) }
\item{DNN Models: M1 (CNN), M2 (LSTM), M3 (BLSTM), M4 (BLSTM with attention) }
\end{itemize}

\subsection{Effect of Oversampling Bullying Instances}
The training datasets had a major problem of class imbalance with posts marked as bullying in the minority. As a result, all models were biased towards labeling the posts as non-bullying. To remove this bias, we oversampled the data from bullying class thrice. That is, we replicated bullying posts thrice in the training data. This significantly improved the performance of all DNN models with major leap in all three evaluation measures. Table \ref{table:oversampling} shows the effect of oversampling for a variety of word embedding methods with BLSTM Attention as the detection model. Results for other models are similar \cite{moreResults}.  We can notice that oversampled datasets (F+, T+, W+) have far better performance than their counterparts (F, T, W respectively). Oversampling particularly helps the smallest dataset Formspring where number of training instances for bullying class is quite small (825) as compared to other two datasets (about 5K and 13K). We also experimented with varying the replication rate for bullying posts \cite{moreResults}. However, we observed that for bullying posts, replication rate of three is good enough.

\begin{table*}[tb]
  \caption{Effect of Oversampling Bullying Posts using BLSTM with attention}
    \begin{tabular}{|c|c|c|c|c|c|c|c|c|c|c|}
   
    \hline
    Dataset & Label & \multicolumn{3}{|c|}{P} & \multicolumn{3}{|c|}{R}& \multicolumn{3}{|c|}{F1} \\
     \cline{3-11}
    & & Random & Glove & SSWE & Random & Glove & SSWE & Random & Glove & SSWE \\
    \hline
    F & bully & 0.52 &0.56 &0.63 & 0.40 & 0.49&0.38 &  0.44 & 0.51& 0.47 \\
  \hline  
      F+ & bully & 0.84 &0.85 &0.90 & 0.98 & 0.97& 0.91&  0.90 &0.90 & 0.91\\
  \hline 
  T & racism & 0.67 &0.74 & 0.76& 0.73 & 0.76& 0.77&  0.70  & 0.75& 0.76 \\
  \hline 
  T+ & racism  & 0.94 &0.90 &0.90 & 0.98 &0.95 & 0.96&  0.96 &0.93 &0.93 \\
  \hline 
  T & sexism & 0.65 &0.86 & 0.83& 0.64 &0.52 &0.47 &  0.65  & 0.65& 0.59 \\
  \hline 

   T+ & sexism  & 0.88 &0.95 & 0.88& 0.97 & 0.91&0.92 &  0.93 &0.91 & 0.90\\
  \hline  
  W & attack & 0.77 & 0.81& 0.82& 0.74 &0.67 & 0.68&  0.76  & 0.74& 0.74 \\
  \hline
  W+ & attack  & 0.81 &0.86 &0.87 & 0.91 &0.89&0.86 &  0.88 & 0.88&0.87 \\
  \hline

\end{tabular}
\label{table:oversampling}
\end{table*}

\subsection{Choice of Initial Word Embeddings and Model}
Initial word embeddings decide data representation for DNN models. However during the training, DNN models modify these initial word embeddings to learn task specific word embeddings. We have experimented with three methods to initialize word embeddings. Please refer to Table \ref{table:wordembedding}. This table shows the effect of varying initial word embeddings for multiple DNN models across datasets. We can notice that initial word embeddings do not have a significant effect on cyberbullying detection when oversampling of bullying posts is done (rows corresponding to F+, T+, W+). In the absence of oversampling (rows corresponding to F, T W), there is a gap in performance of simplest (CNN) and most complex (BLSTM with attention) models. However, this gap goes on reducing with the increase in the size of datasets.


\begin{table*}[tb]
\centering
  \caption{Effect of Choosing Initial Word Embedding Method on F1 Score}
    \begin{tabular}{|c|c|c|c|c|c|c|c|c|c|c|c|c|c|}
   
    \hline
    Dataset & Label & \multicolumn{2}{|c|}{Random} & \multicolumn{2}{|c|}{Glove}& \multicolumn{2}{|c|}{SSWE} \\
     \cline{3-14}
    & & M1 & M4 & M1 & M4 & M1 & M4 \\
  \hline
  F  & bully & 0.30 & 0.44 & 0.34 & 0.51 & 0.34 &0.47\\
  \hline
  F+ & bully & 0.91  &0.90 & 0.93 &0.90 & 0.91  &0.91\\
  \hline
   T & racism & 0.68 & 0.70 & 0.73 & 0.75  & 0.70 &  0.76 \\
   \hline
T+ & racism & 0.90 & 0.96 & 0.95 &  0.93 & 0.93 & 0.93  \\   
\hline
  T & sexism& 0.59 & 0.65 & 0.61 &0.65 & 0.63 &0.59\\
  \hline
  
  T+ & sexism&  0.93 &0.93 & 0.93 & 0.91 &  0.92 & 0.90\\
  \hline
W & Attack  & 0.72 &0.76 & 0.72 &0.74  & 0.74 & 0.74\\
\hline
  W+ & Attack& 0.83  &0.88 & 0.89 & 0.88& 0.88  &0.87\\
  \hline
  \end{tabular}
\label{table:wordembedding}
\end{table*}

Table~\ref{table:modelchoice} compares the performance of four DNN models for three evaluation measures while using SSWE as the initial word embeddings. We have noticed that most of the time LSTM performs weaker than other three models. However, performance gap in the other three models is not significant.


\begin{table*}[tb]
\centering
  \caption{Performance Comparison of Various DNN Models}
  \label{table:modelchoice}
\begin{tabular}{|c|c|c|c|c|c|c|c|c|c|c|c|c|c|}
    \hline
    Dataset & Label &\multicolumn{4}{|c|}{P} &\multicolumn{4}{|c|}{R} &\multicolumn{4}{|c|}{F1}\\
    \cline{3-14}
     &  & M1 & M2 & M3 & M4 & M1 & M2 & M3 & M4& M1 & M2 & M3 & M4\\
    \hline
    F+ & bully &0.93 & 0.91& 0.91&0.90&0.90 &0.85 &0.81 &0.91 &0.91 &0.88 &0.86 &0.91 \\
    \hline
       T+ & racism &0.93 & 0.91& 0.92&0.90 & 0.94 & 0.80& 0.95&0.96 & 0.93 & 0.85& 0.93& 0.93   \\

    	& sexism & 0.92 &0.84 & 0.88&0.88 & 0.92 & 0.93&0.94 & 0.92& 0.92 &0.88 &0.92 & 0.90\\
      \hline    
 W+  & Attack & 0.92 &  0.70&0.90  & 0.87& 0.83 &0.54 &  0.81& 0.86& 0.88 &0.61 & 0.85  &0.87\\	
  \hline
\end{tabular}
\end{table*}

\subsection{Task Specific Word Embeddings}
\label{subsec:taskword}
DNN models learn word embeddings over the training data. These learned embeddings across multiple datasets show the difference in nature and style of bullying across cyberbullying topics and SMPs. Here we report results for BLSTM with attention model. Results for other models are similar. We first verify that important words for each topic of cyberbullying form clusters in the learned embeddings. To enable the visualization of grouping, we reduced dimensionality with t-SNE \cite{maaten08}, a well-known technique for dimensionality reduction particularly well suited for visualization of high dimensional datasets. Please refer to Table~\ref{tab:word clusters}. This table shows important clusters observed in t-SNE projection of learned word embeddings. Each cluster shows that words most relevant to a particular topic of bullying form cluster.

\begin{table} [tb]
  \caption{Embeddings learned using DNNs}
  \label{tab:word clusters}
  \begin{tabular}{|c|p{9cm}|}
     \hline
    Bullying form & Observed Cluster\\
    \hline
	Sexism & kitchen, feminist, feminists, its, feminism, girl, rights, two, female, bitch, head, sexist, woman, girls, blondes, rape \\
    \hline
	Racism & pedophile, murdered, either, israel, mohammed, slave, prophet, muslims, quran, may, islam, religion, war, pay \\
   \hline
	Personal Attack & fuck, fucking, u, little, you, shit, style, faggot, ass, off, changes, suck, see, hate, know, nigger, moron, site \\
   \hline
\end{tabular}
\end{table}

We also observed changes in the meanings of the words across topics of cyberbullying. Table~\ref{tab:meaningchange} shows most similar words for a given query word for two datasets. Twitter dataset which is heavy on sexism and racism, considers word slave as similar to targets of racism and sexism. However, Wikipedia dataset that is about personal attacks does not show such bias.

\begin{table}[tb]
  \caption{Most similar words to the query word across platform }
  \label{tab:meaningchange}
  \centering
  \begin{tabular}{|c|p{4.50cm}|p{4.50cm}|}
     \hline
    Query word &\multicolumn{2}{p{9cm}}{Similar words}\\
    \cline{2-3}
     & Twitter & Wiki \\
    \hline
    slave & feminists, religion, jews, islam, muslims, christians&   sucks, bad, blocked, tried, cannot, can't, didn't, never \\
   \hline
   evidence & god, opinion, eliminated, opinions, murdered, racist, raped&  interested, suggest, yes, love, good, happy, quote, note, useful\\
    \hline
    fat & female, woman, face, women, kids, fan, blonde, friends & blocked, sorry, bad, used, tried, cannot, banned, never, fuck \\
    \hline
    gay &  die, ask, fake, child, babies, females, wife, female, woman &  bad, sorry, used, blocked, tried, fuck, fucking, that's, notice, shit\\
   \hline
\end{tabular}
\end{table}

\subsection{Transfer Learning}
We used transfer learning to check if the knowledge gained by DNN models on one dataset can be used to improve cyberbullying detection performance on other datasets. We report results where BLSTM with attention is used as the DNN model. Results for other models are similar \cite{moreResults}. We experimented with following three flavors of transfer learning.

\underline{Complete Transfer Learning (TL1)}: In this flavor, a model trained on one dataset was directly used to detect cyberbullying in other datasets without any extra training. TL1 resulted in significantly low recall indicating that three datasets have different nature of cyberbullying with low overlap (Table ~\ref{table:transfer}). However precision was relatively higher for TL1, indicating that DNN models are cautious in labeling a post as bully (Table ~\ref{table:transfer}). TL1 also helps to measure similarity in nature of cyberbullying across three datasets. We can observe that bullying nature in Formspring and Wikipedia datasets is more similar to each other than the Twitter dataset. This can be inferred from the fact that with TL1, cyberbullying detection performance for Formspring dataset is higher when base model is Wikipedia (precision =0.51 and recall=0.66)as compared to Twitter as the base model (precision=0.38 and recall=0.04). Similarly, for Wikipedia dataset, Formspring acts as a better base model than Twitter while using TL1 flavor of transfer learning. Nature of SMP might be a factor behind this similarity in nature of cyberbullying. Both Formspring and Wikipedia are task oriented social networks (Q\&A and collaborative knowledge repository respectively) that allow anonymity and larger posts. Whereas communication on Twitter is short, free of anonymity and not oriented towards a particular task.

\underline{Feature Level Transfer Learning (TL2)}: In this flavor, a model was trained on one dataset and only learned word embeddings were transferred to another dataset for training a new model. As compared to TL1, recall score improved dramatically with TL2 (Table ~\ref{table:transfer}). Improvement in precision was also significant (Table ~\ref{table:transfer}). These improvements indicate that  learned word embeddings are an essential part of knowledge transfer across datasets for cyberbullying detection.

\underline{Model Level Transfer Learning (TL3)}: In this flavor, a model was trained on one dataset and learned word embeddings, as well as network weights, were transferred to another dataset for training a new model. TL3 does not result in any significant improvement over TL2. This lack of improvement indicates that transfer of network weights is not essential for cyberbullying detection and learned word embeddings is the key knowledge gained by the DNN models.

DNN based models coupled with transfer learning beat the best-known results for all three datasets. Previous best F1 scores for Wikipedia \cite{wulczyn17} and Twitter \cite{badjatiya17} datasets  were 0.68 and 0.93 respectively. We achieve F1 scores of 0.94 for both these datasets using BLSTM with attention and feature level transfer learning (Table ~\ref{table:transfer}). For Formspring dataset, authors have not reported F1 score. Their method has accuracy score of 78.5\%  \cite{reynolds11}. We achieve F1 score of 0.95 with accuracy score of 98\% for the same dataset. 


\begin{table*} [tb]
\centering
  \caption{Comparison of Transfer Learning Methods Using Precision }
  \begin{tabular}{|c|r|c|c|c|c|c|c|c|c|c|}
    \hline
Metric & \backslashbox{ Test }{ Train } & \multicolumn{3}{|c|}{F+} & \multicolumn{3}{|c|}{T+}& \multicolumn{3}{|c|}{W+} \\
       \cline{3-11}
    & &  TL1 & TL2 & TL3 &  TL1 & TL2 & TL3 &  TL1 & TL2 & TL3 \\
  \hline
 \multirow{3}{*}{Precision} & F &   - & - & - &0.38 & 0.90& 0.88& 0.51 & 0.92 & 0.85\\
  \cline{2-11}
   & T & 0.83 & 0.88 & 0.90& - & - & - & 0.72 &0.91 & 0.90 \\
 \cline{2-11}
  & W & 0.82 & 0.92 & 0.91& 0.68 & 0.90& 0.91& - &- & -\\
  \hline
  \multirow{3}{*}{Recall} & F & -& - &- &0.04 & 0.98 & 0.98 & 0.66 &0.98 &0.99\\
  \cline{2-11} 
 & T &0.01 & 0.99 &0.99 & - & - & - & 0.17 &0.98 & 0.99\\
  \cline{2-11}
 & W &0.21& 0.96 &0.96 & 0.05 &0.97 &0.96 & -& - & -\\
  \hline
 \multirow{3}{*}{F1-score} & F & -& - & -&0.07& 0.95 &0.93 &0.58 &0.95 &0.92 \\
  \cline{2-11} 
 &  T & 0.03&0.93 & 0.94& -& - & -&0.28& 0.94& 0.94\\
  \cline{2-11}
 &  W &0.35& 0.94 & 0.94&0.10 & 0.94& 0.94& - &- & - \\
  \hline  
  
  \end{tabular}
\label{table:transfer}
\end{table*}

\subsection{Conclusion and Future Work}
We have shown that DNN models can be used for cyberbullying detection on various topics across multiple SMPs using three datasets and four DNN models. These models coupled with transfer learning beat state of the art results for all three datasets. These models can be further improved with extra data such as information about the profile and social graph of users. Most of the current datasets do not provide any information about the severity of bullying. If such fine-grained information is made available, then cyberbullying detection models can be further improved to take a variety of actions depending on the perceived seriousness of the posts.

\bibliographystyle{abbrv}
\bibliography{sigproc} 

\begin{landscape}
\begin{table}
  \caption{Summary of Related Work}
  \label{tab:papers}
  \begin{tabular}{|p{0.8cm}|p{0.8cm}|p{1.7cm}|p{3.1cm}|p{2.7cm}|p{2cm}|p{1.8cm}|p{1.2cm}|p{1.3cm}|p{2.9cm}|}
     \hline
    Paper & Year & Target SMPs & Data Features & Model Used & Cyberbullying Topics & Bottlenecks & Dataset size & Metric Used & Metric Value\\
    \hline
     \cite{dinakar11} & 2011 & Youtube &  TF-IDF, list of swear words etc. & Naive bayes, SVM, J48, JRip & Sexuality, Race and culture, Intelligence & B1, B3 & 4500 & Accuracy & Sexuality(80.20\%) Intelligence(70.39\%) Race(68.30\%)\\
   \hline
   \cite{reynolds11} & 2011 & Formspring & number of "bad" words
(NUM), density of "bad" words (NORM) & SMO, IBK, J48, JRip & Bully/ Not bully & B1, B2, B3 & 3915 & Accuracy & 78.5\% \\
   \hline
   \cite{djuric15} & 2015& Yahoo & distributed representations of comments(paragraph2vec) & continuous BOW (CBOW) & Hate Speech/ Clean & B1, B2 & \textasciitilde951k  & AUC & 0.80\\
   \hline
   \cite{van15} & 2015 & Ask.fm & word unigram and bigram bags-of-words, character trigram bag-of-words, sentiment lexicon features(comment2vec) & SVM & Threat, Insult, Defense, Sexual Talk, Defamation, Encouragements
and Swear & B1, B3 &  \textasciitilde91k &  F1-score & 55.39\\
   \hline
\cite{nobata16} & 2016 & Yahoo & word and character N-grams, Linguistic features, Syntactic and Distributional Semantics &  Vowpal Wabbits
regression model & Abusive/ Non Abusive & B1, B2, B3 & Finance (\textasciitilde759k), News (\textasciitilde1390k)  & AUC & 0.90\\
\hline
   \cite{badjatiya17}  & 2017& Twitter  & TF-IDF values, BoWV over Global Vectors, task-specific embeddings & FastText, CNNs, LSTMs, GBDT & Sexism, Racism, None & B1 & 16k & F1-score & 0.93\\
   \hline
   \cite{wulczyn17}  & 2017& Wikipedia & char-n grams, word n-grams & LR, MLP & Personal Attack & B1, B2 & 100k & AUC & 96.59 \\

   \hline
\end{tabular}
\end{table}
\end{landscape}

\end{document}